\documentclass{PoS}
\usepackage{multirow, booktabs}
\usepackage{siunitx}
\usepackage{url}
\usepackage{graphicx}

\usepackage{amssymb}
\usepackage{fontenc}
\usepackage{times}
\usepackage{mathptmx}
\usepackage{graphicx}

\graphicspath{ {/./}}
\DeclareGraphicsExtensions{.png,.pdf}
\title{First Results of Eta Car Observations with H.E.S.S. II}

\ShortTitle{First Results of Eta Car Observations with H.E.S.S. II}

\author{\speaker{E. Leser} $^{1}$,  
S. Ohm $^{2}$, 
M. F\"u\ss{}ling $^{2}$, 
M. de Naurois $^{3}$,  
K. Egberts $^{1}$, 
P. Bordas $^{4}$, 
S. Klepser $^{2}$, 
O. Reimer $^{5}$, 
A. Reimer $^{5}$, 
J. Hinton $^{4}$, 
for the H.E.S.S. collaboration\\
$^{1}$ Institut f\"ur Physik und Astronomie, Universit\"at Potsdam, Karl-Liebknecht-Strasse 24/25, D 14476 Potsdam, Germany \\
$^{2}$ DESY, D-15738 Zeuthen, Germany\\
$^{3}$ Laboratoire Leprince-Ringuet, Ecole Polytechnique, CNRS/IN2P3, F-91128 Palaiseau, France\\
$^{4}$ Max-Planck-Institut f\"ur Kernphysik, Heidelberg, Germany\\
$^{5}$ Institut f\"ur Astro- und Teilchenphysik, Leopold-Franzens-Universit\"at Innsbruck\\
E-mail: \email{contact.hess@hess-experiment.eu}}

\abstract{Eta Carinae (or Eta Car) is a colliding-wind binary that shows non-thermal emission from hard X-rays to high-energy $\gamma$-rays. The $\gamma$-ray spectrum exhibits two spectral components, where the high-energy component extends up to 300 GeV.
Previous observations of Eta Carinae with the High Energy Stereoscopic System (H.E.S.S.) resulted in upper limits. With the addition of the large 28-m central telescope to the H.E.S.S. array in 2012 the lower bound on the energy range has significantly been reduced. This lowers the energy threshold of the analysis compared to the published results due to the improved instrument sensitivity at energies below 400 GeV. Eta Carinae has been regularly observed in the following years with H.E.S.S. II.
Here we report on the first results of Eta Carinae observations, which also cover the periastron passage in 2014 - the time of maximum $\gamma$-ray emission seen in GeV $\gamma$-rays with Fermi-LAT.}

\FullConference{35th International Cosmic Ray Conference\\
		 10-20 July, 2017\\
		 Bexco, Busan, Korea}

\begin{document}
\section*{Detection of Eta Carinae with H.E.S.S.}
Eta Carinae, member star of open stellar cluster Trumpler 16, is embedded in the complex HII region Carina Nebula which is host to eight stellar clusters. 
The binary stellar system of Eta Carinae, located at a distance of 2.3 kpc in the Carina-Sagittarius Arm of the Milky Way, is composed of a luminous blue variable ($\textrm{m}_{1} = 100\,\textrm{M}_\odot$) and its lighter ($\textrm{m}_{2} = 30\,\textrm{M}_\odot$) companion star of O- or B-type \cite{Smith2006}. Its - very eccentric ($\textrm{e}\,\sim\,0.9$) - orbit spans a period of 5.5 years and the last periastron passage took place in summer 2014 \cite{Damineli:2007tc}, \cite{Nielsen:2007ht}. 

Eta Carinae has experienced from massive eruptions. An outburst known as the Great Eruption happened in the 1840s and expelled $\sim 15\,\textrm{M}_{\odot}$ forming the Homunculus Nebula. Another smaller nebula, the little Homunculus Nebula, was created by a following outburst. Although very well studied from radio to X-ray wavelengths the binary system has only recently been observed over a full orbit in $\gamma$-rays by the \textit{Fermi} satellite following its detection in 2009. This makes Eta Carinae the only colliding-wind binary detected in high-energy $\gamma$-rays. Both stars drive supersonic winds. The mass outflow of the primary is $\dot{m_1} \approx 2\, \textrm{x}\, 10^{-4} \,\textrm{M}_\odot \, \textrm{yr}^{-1}$ and $\dot{m_2} \approx 2 \,\textrm{x}\,10^{-5} \,\textrm{M}_\odot \,\textrm{yr}^{-1}$ of the companion star \cite{Hillier2001ApJ}, \cite{Parkin2009}. The companion star has the thinner and faster stellar wind with $v_2 \approx\,3000\,\textrm{km}\,\textrm{s}^{-1}$, compared to the speed of the wind of the primary star $v_1 \approx\,600\,\textrm{km}\,\textrm{s}^{-1}$ \cite{PittardCorcoran2002}. Both winds together provide a total kinetic energy in the order of a few 10$^{37}\,\mathrm{erg}\,\mathrm{s}^{-1}$\cite{Smith2006}. 

Following \cite{EichlerUsov1993} and \cite{Reimer2005} particles can be accelerated to non-thermal energies in shock regions that form in the collision of the winds. Protons accelerated in the shocks could interact with nuclei giving rise to the production of neutral pions and their subsequent decay into gamma rays with MeV - GeV energies. Electrons could also contribute to the emission through inverse Compton scattering. 

\begin{figure}
\centering
\includegraphics[width=0.45\textwidth]{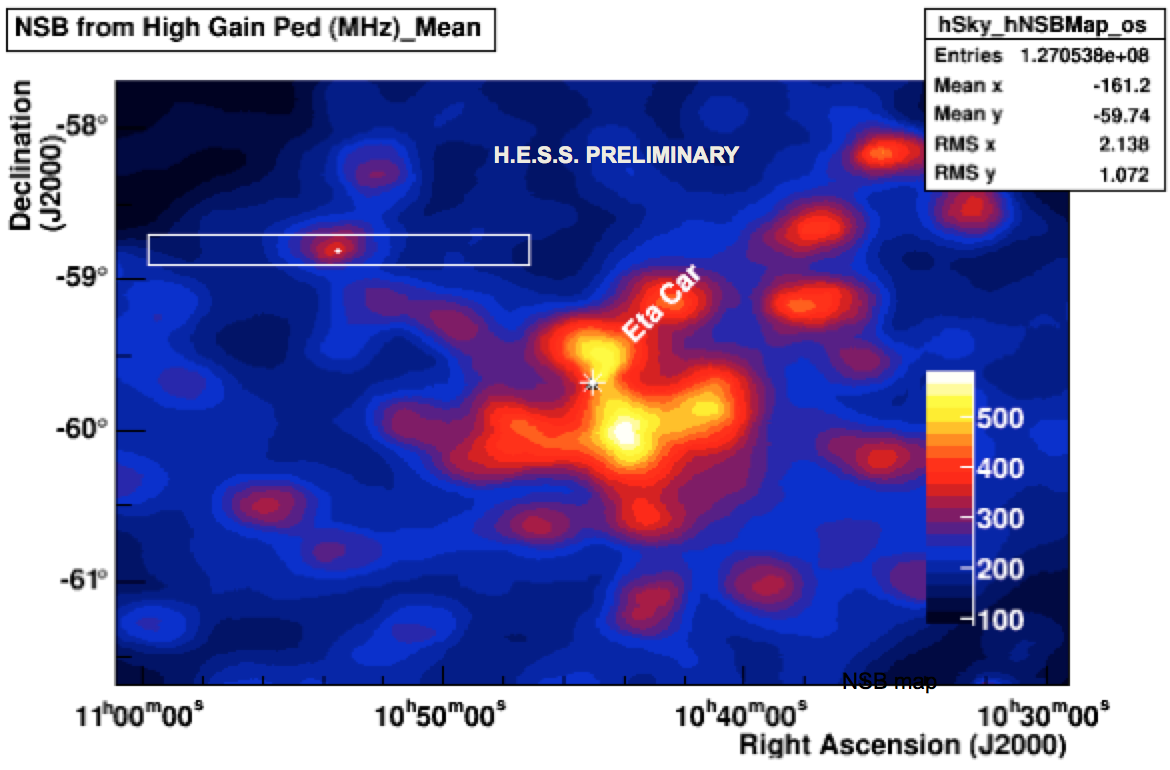}
\caption{\label{fig:NSB}Night sky background in MHz. The box in the upper left depicts a region where camera pixels had to be turned off because of too high NSB rates. This region was used to test how much contribution in terms of excess can be exptected from the highest NSB regions. }
\end{figure}

The presence of a high-energy component in the $\gamma$-ray emission of Eta Carinae detected by \textit{Fermi} motivated the H.E.S.S. collaboration to search for emission in the transition region in energy between the \textit{Fermi} and H.E.S.S. instruments. In 2012 the results of this search were published in a paper with upper limits \cite{HESS2012EtaCar}. Back then the H.E.S.S. array consisted of four imaging atmospheric Cherenkov telescopes with 12 m diameter dishes situated in Khomas Highland, Namibia \cite{HESS2006Crab}. The energy threshold of 470 GeV prevented to cover the energy range where the \textit{Fermi} measurements ceased. 
Lowering the energy threshold of the H.E.S.S. array was a driving factor behind the upgrade that took place in 2012 when a fifth telescope equipped with a big dish of 28 m in diameter was added. The threshold could be decreased to tens of GeV. The observations were taken in 2014 and 2015, right before periastron passage and shortly after (phases 0.78 - 1.1). A part of the data set (after periastron passage) was obtained using only the biggest, 28m diameter telescope, while before periastron there is also data with all five telescopes available. In total 25 hours of observations were taken. Compared to other objects in the H.E.S.S. field of view the night sky background (NSB) around Eta Carinae is very strong (10 times higher than the average) and also inhomogeneous. In order to carry out observations it was necessary to raise the threshold in the cameras collecting the Cherenkov photons. This way it could be ensured that the instrument does not trigger on NSB photons. Still there is the possibility that NSB photons contaminate the real Cherenkov light seen from the air shower. Several systematic tests were applied to assess the possible impact of the NSB on the analysis. At the lowest level of data processing a cleaning algorithm was used that discards 10 \% of the noisiest camera pixels and after a cut on 3 times the measured NSB rate only 1.5 \% of pure NSB photons survive. The distributions of Cherenkov showers in the camera were checked for homogeneity to avert single bright pixels from enhancing the Cherenkov shower. The measured NSB counts map shows strong features around the position Eta Carinae as expected but the maximum is seen in a region south of the colliding wind binary (compare figure \ref{fig:NSB}). 
In case some NSB photons remain these are treated by using Monte Carlo simulations that exactly resemble the observational conditions including the measured NSB at every pointing position of the telescopes. Two independent analysis chains were used to analyse the obtained (monoscopic and stereoscopic) data, shown here \cite{Holler2015} and \cite{Murach2015} to check the results. They coincidently report a significant emission from the direction of Eta Carinae. The significance is 13.6 $\sigma$ pre-trial for the combined data set of 25 hours. The energy threshold obtained is $\sim $200$\,$GeV which is 60\% lower compared to the threshold in the 2012  H.E.S.S. paper. In 25 hours the H.E.S.S. instrument measured 1070 $\gamma$-ray candidate events. Details will be presented in a forthcoming publication by the H.E.S.S. Collaboration (in preparation).
\begin{figure}
\centering
\includegraphics[width=0.4\textwidth]{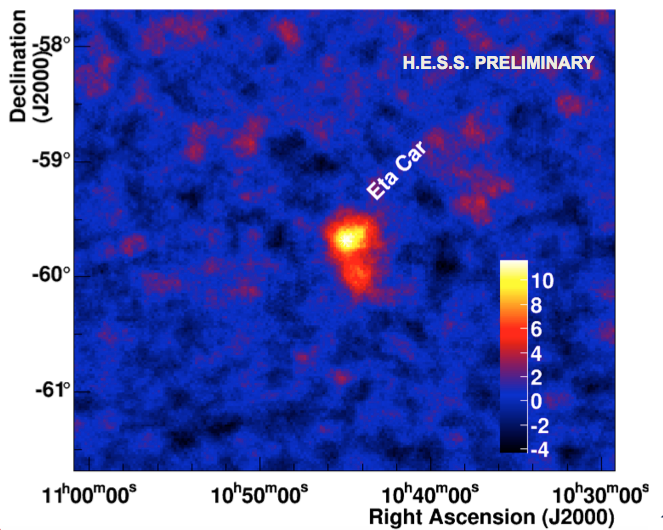}
\caption{\label{fig:Sign}Significance map for 25 hours of monoscopic observations with the biggest telescope using an oversampling radius of 0.1$\,^{\circ}$.}
\end{figure}

\newpage
\section*{Acknowledgments}
The support of the Namibian authorities and of the University of Namibia in facilitating the construction and operation of H.E.S.S. is gratefully acknowledged,  as is the support by the German Ministry for Education and Research (BMBF), the Max Planck Society, the German Research Foundation (DFG), the French Ministry for Research, the CNRS-IN2P3 and the Astroparticle Interdisciplinary Programme of the CNRS, the U.K. Science and Technology Facilities Council (STFC), the IPNP of the Charles University, the Czech Science Foundation,  the Polish Ministry of Science and Higher Education, the South African Department of Science and Technology and National Research Foundation, the University of Namibia, the Innsbruck University,  the Austrian Science Fund (FWF), and the Austrian Federal Ministry for Science, Research and Economy, and by the University of Adelaide and the Australian Research Council. We appreciate the excellent work of the technical support staff in Berlin, Durham, Hamburg, Heidelberg, Palaiseau, Paris, Saclay, and in Namibia in the construction and operation of the equipment. This work benefitted from services provided by the H.E.S.S. Virtual Organisation, supported by the national resource providers of the EGI Federation.

\newpage
\bibliographystyle{apalike} 
\bibliography{BibStatus.bib}


\end{document}